\documentclass[pdflatex,sn-mathphys-num]{sn-jnl}

\usepackage{graphicx}%
\usepackage{multirow}%
\usepackage{amsmath,amssymb,amsfonts}%
\usepackage{amsthm}%
\usepackage{mathrsfs}%
\usepackage[title]{appendix}%
\usepackage{xcolor}%
\usepackage{textcomp}%
\usepackage{manyfoot}%
\usepackage{booktabs}%
\usepackage{algorithm}%
\usepackage{algorithmicx}%
\usepackage{algpseudocode}%
\usepackage{listings}%

\usepackage{bm}
\usepackage{longtable}
\usepackage{array}
\usepackage{enumitem}

\hypersetup{hidelinks}

\providecommand{\ket}[1]{\ensuremath{\left|#1\right\rangle}}
\providecommand{\bra}[1]{\ensuremath{\left\langle#1\right|}}

\theoremstyle{thmstyleone}%
\theoremstyle{thmstyletwo}%

\theoremstyle{thmstylethree}%

\begin{document}

\title[QAOA for Layered Architecture Recovery]{QAOA for Layered Architecture Recovery: Feasibility, Limits, and the Role of Decomposition}


\author[1]{\fnm{Julien} \sur{Cardinal}}\email{julien.cardinal.1@ens.etsmtl.ca}

\author[1]{\fnm{Andrew} \sur{Chedid}}\email{andrew.chedid.1@ens.etsmtl.ca}

\author*[1]{\fnm{Imen} \sur{Benzarti}}\email{imen.benzarti@etsmtl.ca}

\author[1]{\fnm{Ghizlane} \sur{El Boussaidi}}\email{ghizlane.elboussaidi@etsmtl.ca}

\author[1]{\fnm{Christophe} \sur{P\`ere}}\email{christophe.pere@etsmtl.ca}

\author[1]{\fnm{Jean-Marc} \sur{Robert}}\email{jean-marc.robert@etsmtl.ca}

\affil[1]{\orgdiv{Department of Software Engineering and IT},
\orgname{\'Ecole de technologie sup\'erieure (\'ETS), Universit\'e du Qu\'ebec}, \orgaddress{\street{1100 Notre-Dame St.\ W.}, \city{Montr\'eal}, \postcode{H3C 1K3}, \state{Qu\'ebec}, \country{Canada}}}

\abstract{Layered architecture recovery (LAR) reconstructs a system's architectural layers from its package dependencies, and is defined as a combinatorial optimisation problem solved with classical search. We investigate whether the Quantum Approximate Optimization Algorithm (QAOA) can be applied to LAR, and with what practical limitations. We formulate LAR as a Quadratic Semi-Assignment Problem, transform it into a Quadratic Unconstrained Binary Optimisation (QUBO) problem in which each package is penalised unless it is assigned to exactly one
layer, and encode it as an Ising Hamiltonian solved by QAOA. To reach instances beyond the qubit budget of a single circuit, we also apply an impact-guided decomposition that solves the problem as a sequence of smaller sub-problems. We evaluate the approach on three real Java systems ($15$, $18$, and $24$ qubits), studying the effect of circuit depth and sampling budget (RQ1), a comparison with a classical Tabu Search baseline (RQ2), robustness to simulated quantum noise (RQ3), and the decomposition (RQ4). We find that monolithic QAOA is feasible and accurate on the two smaller systems but fails on the $24$-qubit instance, that the sampling budget rather than circuit depth governs its accuracy, and that it is competitive with but does not outperform Tabu Search. Decomposition restores reliable recovery on the $24$-qubit instance, raising the optimal rate from $6.7\%$ to $93.3\%$; its accuracy is governed by the size of each sub-problem and, unlike the monolithic pipeline, improves as the sampling budget is reduced. Our results indicate that, in the current noisy intermediate-scale regime, problem decomposition, rather than larger circuits or more sampling, is the key to applying QAOA to software architecture recovery. }


\maketitle

\section{Introduction}
\label{sec:intro}

Understanding a software system's architecture is essential for its maintenance and evolution~\cite{ducasse2009}. As software systems evolve, however, their implementations often drift away from the intended architectural design. This phenomenon is referred to as \emph{architecture erosion}~\cite{desilva2012,li2022erosion}. Architectural decisions become implicit, structural dependencies accumulate, and the original organization of the system may no longer be directly observable from the source code~\cite{ducasse2009}. Architecture recovery aims to reconstruct this architectural structure from implementation artifacts, by supporting program comprehension, software evolution, and the identification of architectural degradation~\cite{ducasse2009,li2022erosion}.

Layered Architecture Recovery (LAR), as formalised by Belle et al.~\cite{belle2015}, addresses this problem by assigning software packages to architectural layers based on their observed dependencies. A good assignment favours dependencies toward the immediately adjacent lower layer while penalising architectural violations such as back-calls, skip-calls, and intra-layer dependencies~\cite{belle2015}. Belle et al. formulate LAR as a Quadratic Semi-Assignment Problem (QSAP), a computationally difficult combinatorial optimization problem related to the Quadratic Assignment Problem (QAP)~\cite{belle2015,pardalos1994}. Architecture recovery and software remodularization problems of this kind are commonly addressed using heuristic and metaheuristic search techniques~\cite{belle2015,praditwong2011,mkaouer2015}; in particular, Belle et al. employ Tabu Search to solve the LAR formulation. As the number of packages and candidate layers increases, however, the number of possible assignments grows combinatorially, making the exploration of the solution space increasingly challenging~\cite{pardalos1994,belle2015} and motivating the investigation of alternative optimization paradigms.

Quantum optimization provides one such alternative. In particular, the Quantum Approximate Optimization Algorithm (QAOA)~\cite{farhi2014} is a hybrid quantum--classical variational algorithm designed for combinatorial optimization. QAOA encodes an optimization objective into a problem Hamiltonian and alternates problem and mixer operators in a parameterised quantum circuit, whose parameters are optimized through a classical routine. QAOA is particularly relevant to optimization problems that can be expressed as Quadratic Unconstrained Binary Optimization (QUBO) models. Within software engineering, quantum and quantum-inspired optimization has received increasing attention, although recent work remains concentrated in a limited number of problem domains, particularly software testing~\cite{wang2024,trovato2025,wang2025roadmap,zhang2026survey}. Applications include test-case optimization~\cite{wang2024}, test-case minimization~\cite{wang2024minimization}, and test-case selection~\cite{trovato2025}, while other software engineering activities remain comparatively underexplored~\cite{zhang2026survey}.

LAR is particularly interesting from a quantum-optimization perspective because its QSAP formulation can be transformed into a QUBO and subsequently encoded as an Ising Hamiltonian suitable for QAOA. At the same time, the characteristics of current quantum computing raise important questions about whether such a formulation can produce useful solutions in practice. \textbf{To the best of our knowledge, QAOA has not previously been investigated for layered architecture recovery.}

In this work, we investigate the \emph{feasibility, effectiveness, and practical limitations} of applying QAOA to LAR. Rather than seeking to demonstrate quantum advantage, which would be premature at the scale of current quantum hardware, our objective is to determine how LAR can be formulated for QAOA and to empirically characterise the behaviour of the resulting hybrid quantum--classical approach. In particular, we examine how circuit depth and sampling budget affect solution quality, how QAOA compares with a classical Tabu Search baseline, and how simulated quantum noise affects its behaviour. Experiments are conducted using ideal and noisy quantum simulators with configurations intended to reflect the resource constraints of near-term quantum execution.

This study builds on our earlier experience report on migrating a QAOA implementation across Qiskit versions~\cite{cardinal2025}. That work focused primarily on the engineering and validation of the QAOA implementation, comparing a custom pipeline with the Qiskit Algorithms library and highlighting the importance of the sampling budget for solution quality. The present work moves from implementation validation to a systematic empirical investigation of QAOA for LAR. It introduces the problem formulation and evaluates the approach across multiple software systems, parameter configurations, a classical baseline, and noisy execution environments.

The main contributions of this paper are: (1) A formulation of Layered Architecture Recovery for quantum optimisation, transforming its QSAP representation into a QUBO with one-hot constraint penalties and subsequently into an Ising Hamiltonian suitable for QAOA, (2) A hybrid quantum--classical QAOA approach for LAR that combines parameterised quantum-circuit construction, COBYLA-based parameter optimisation, measurement, classical post-processing of sampled candidate solutions, and validation against the LAR assignment constraints, (3) An impact-guided decomposition of the LAR QUBO that bounds the width of each quantum circuit to an active set of packages, allowing instances that exceed the qubit budget of a single circuit to be recovered as a sequence of smaller sub-problems, (4) An empirical evaluation on three real-world Java systems investigating the sensitivity of QAOA to circuit depth and sampling budget (RQ1), its effectiveness and computational effort relative to a classical Tabu Search baseline (RQ2), its robustness under simulated quantum noise (RQ3), and the effect of impact-guided decomposition on recovery quality and feasibility (RQ4). (5) Empirical insights into the practical limitations of QAOA for software architecture recovery on resource-constrained quantum systems, including the roles of sampling, classical post-processing, problem size, decomposition, and noise.

This article is an extended version of our paper presented at the 7th International Workshop on Quantum Software Engineering (Q-SE 2026). It substantially extends the workshop version with : the full QSAP-to-QUBO-to-Ising formulation, the impact-guided decomposition and RQ4, the Tabu Search comparison, the noise study, and  a complete re-evaluation on three systems.

The remainder of the paper is organised as follows. Section~\ref{sec:background} introduces Layered Architecture Recovery, QAOA, and impact-guided decomposition. Section~\ref{sec:formulation} presents the QSAP formulation of LAR and its transformation into QUBO and Ising representations. Section~\ref{sec:approach} describes the proposed QAOA-based recovery approach. Section~\ref{sec:eval} presents the experimental design and research questions, and Section~\ref{sec:results} reports the results. Sections~\ref{sec:discussion}--\ref{sec:related} discuss the findings, threats to validity, and related work. Finally, Section~\ref{sec:conclusion} concludes the paper and outlines directions for future work.

\section{Background}
\label{sec:background}

\subsection{Layered Architecture Recovery}
\label{sec:bg-lar}

We consider the Layered Architecture Recovery (LAR) problem introduced by Belle et al.~\cite{belle2015}. Given a software system and its package-dependency graph, LAR seeks an assignment of each package to exactly one architectural layer such that the resulting structure conforms as closely as possible to the layered architectural style. In a well-formed layered architecture, dependencies preferably flow downward toward the immediately adjacent lower layer, whereas dependencies that violate this organization incur a penalty.

Belle et al.~\cite{belle2015} distinguish three types of architectural violations:
\begin{itemize}[leftmargin=1.4em,itemsep=1pt]
\item \emph{Back-calls}: a package depends on a package located in a higher layer. These constitute the strongest violation of the intended dependency direction.
\item \emph{Skip-calls}: a package depends on a package located more than one layer below it. These bypass an intermediate layer and are treated as less severe violations than back-calls.
\item \emph{Intra-layer dependencies}: a package depends on another package assigned to the same layer.
\end{itemize}

Architecture recovery can therefore be formulated as an optimization problem: among the possible package-to-layer assignments, the objective is to find an assignment that minimizes the total penalty associated with these dependency violations. Belle et al.~\cite{belle2015} formulate this problem as a Quadratic Semi-Assignment Problem (QSAP). Section~\ref{sec:formulation} presents this formulation and its transformation into a representation suitable for quantum optimization.

\subsection{The Quantum Approximate Optimization Algorithm}
\label{sec:bg-qaoa}

The Quantum Approximate Optimization Algorithm (QAOA)~\cite{farhi2014} is a hybrid quantum--classical variational algorithm designed for combinatorial optimization. It encodes the objective function of an optimization problem in a problem Hamiltonian $H_C$ and constructs a parameterized quantum state by alternating a problem operator with a mixer operator. A classical optimizer then adjusts the circuit parameters to minimize the expectation value of the problem Hamiltonian~\cite{cerezo2021}.

Starting from the equal-superposition state $\ket{+}^{\otimes N}$ over $N$ qubits, a depth-$p$ QAOA circuit applies $p$ alternating layers of the problem and mixer unitaries
\begin{equation}
U_C(\gamma) = e^{-i\gamma H_C}, \qquad
U_M(\beta) = e^{-i\beta H_M},
\end{equation}
where $H_C$ encodes the optimization objective and $H_M$ promotes exploration of the solution space. The resulting state is
\begin{equation}
\ket{\psi(\bm{\gamma},\bm{\beta})}
= U_M(\beta_p)U_C(\gamma_p)\cdots
U_M(\beta_1)U_C(\gamma_1)
\ket{+}^{\otimes N}.
\end{equation}
Each QAOA layer introduces one problem parameter $\gamma$ and one mixer parameter $\beta$; consequently, a depth-$p$ circuit contains $2p$ variational parameters.

The objective optimized by the classical routine is the expectation value
\begin{equation}
F_p(\bm{\gamma},\bm{\beta})
= \bra{\psi(\bm{\gamma},\bm{\beta})}
H_C
\ket{\psi(\bm{\gamma},\bm{\beta})},
\end{equation}
whose minimizing parameters are
\begin{equation}
(\bm{\gamma}^{\ast},\bm{\beta}^{\ast})
= \arg\min_{\bm{\gamma},\bm{\beta}}
F_p(\bm{\gamma},\bm{\beta}).
\end{equation}

Once the parameters have been optimized, the circuit is measured repeatedly. Each execution, or \emph{shot}, produces a bitstring representing a candidate solution, and multiple shots yield an empirical probability distribution over the sampled solutions. The number of shots therefore determines the sampling budget available for identifying promising candidate states. Both circuit depth $p$ and the number of shots are important practical parameters of QAOA and are investigated empirically in this study.

Current quantum computers operate in the so-called Noisy Intermediate-Scale Quantum (NISQ) regime~\cite{preskill2018}: they provide a limited number of qubits, lack full error correction, and are subject to gate and measurement noise that accumulates with circuit depth. These constraints bound both the size of the problem that can be encoded, since the qubit count scales with the number of decision variables, and the depth of circuit that can be executed reliably. In practice, they push QAOA toward shallow circuits combined with large shot budgets and classical post-processing of the sampled candidates. The circuit depth $p$, the number of shots, and the robustness of the algorithm to hardware noise are therefore the parameters that most directly govern its practical usefulness, and we study them empirically in Sections~\ref{sec:results} (RQ1 and RQ3).

QAOA is particularly relevant to optimisation problems that can be expressed as Quadratic Unconstrained Binary Optimisation (QUBO) models, which map directly onto the Ising Hamiltonian that QAOA optimises. Section~\ref{sec:formulation} presents this transformation, and Section~\ref{sec:approach} describes how the resulting QAOA circuit is used for architecture recovery.

\subsection{Impact-Guided Decomposition}
\label{sec:bg-decomposition}

Encoding LAR for QAOA requires one qubit per package--layer variable, so the qubit count grows with system size and, under the NISQ constraints noted above, soon exceeds what a single circuit can handle. To address this, we build on the Impact-Guided Decomposition QAOA (IGDec-QAOA) approach of Wang et al.~\cite{wang2024}, originally proposed for test case optimisation, which pairs QAOA with a decomposition strategy adapted from D-Wave's \emph{qbsolv}~\cite{qbsolv} to solve a large QUBO as a sequence of smaller sub-problems. Section~\ref{sec:decomposition} describes how we adapt this strategy to LAR.

\section{Problem Formulation}
\label{sec:formulation}

This section formulates LAR as a QSAP (Section~\ref{sec:qsap}), transforms it into the QUBO and Ising form required by QAOA (Section~\ref{sec:qubo}), and describes the decomposition used to keep the problem within reach of near-term quantum resources (Section~\ref{sec:decomposition}).

\subsection{QSAP Formulation of LAR}
\label{sec:qsap}

Following Belle et al.~\cite{belle2015} and the QAP literature~\cite{pardalos1994}, we model LAR as a Quadratic Semi-Assignment Problem. Let there be $m$ packages, indexed by $i,j \in \{1,\dots,m\}$, and $n$ layers, indexed by $k,l \in \{1,\dots,n\}$. The matrix $W = (w_{ij})$ gives the dependency weight between packages---the number of calls from package $i$ to package $j$---and the matrix $C = (c_{kl})$ gives the cost of placing two dependent packages in layers $k$ and $l$. The binary decision variable $x_{ik}=1$ if package $i$ is assigned to layer $k$, and $0$ otherwise.

The optimization problem is
\begin{align}
\min \quad & \sum_{i=1}^{m}\sum_{j=1}^{m}\sum_{k=1}^{n}\sum_{l=1}^{n}
            w_{ij}\, c_{kl}\, x_{ik}\, x_{jl} \label{eq:qsap-obj}\\
\text{s.t.} \quad & \sum_{k=1}^{n} x_{ik} = 1 && \forall\, i \in \{1,\dots,m\}
            \label{eq:onehot}\\
& x_{ik} \in \{0,1\} && \forall\, i \in \{1,\dots,m\},\; k \in \{1,\dots,n\}.
            \label{eq:binary}
\end{align}
Constraint~\eqref{eq:onehot} enforces that each package is assigned to exactly one layer.

The cost matrix $C$ encodes the penalties of Section~\ref{sec:bg-lar}. For a directed dependency from package $i$ (layer $k$) to package $j$ (layer $l$):
\begin{itemize}[leftmargin=1.4em,itemsep=1pt]
  \item \emph{Adjacent} ($k = l+1$): the desirable case, $c_{kl}=0$;
  \item \emph{Same layer} ($k = l$): intra-layer dependency, $c_{kl}=\gamma$;
  \item \emph{Skip} ($k \ge l+2$): skip-call, $c_{kl}=\alpha$;
  \item \emph{Back} ($k \le l-1$): back-call, $c_{kl}=\beta$.
\end{itemize}
These rules define $C$ for any number of layers $n$. For example, with $n=3$,
\begin{equation}
C =
\begin{pmatrix}
\gamma & \beta  & \beta \\
0      & \gamma & \beta \\
\alpha & 0      & \gamma
\end{pmatrix}.
\end{equation}

\subsection{QUBO and Ising Formulation}
\label{sec:qubo}

To solve LAR on a quantum device, we transform the QSAP into a Quadratic Unconstrained Binary Optimisation (QUBO) problem, the form most directly amenable to QAOA. The transformation follows the standard Ising construction of Lucas~\cite{lucas2014} and the Qiskit Algorithms library~\cite{javadiabhari2024}.

The QUBO matrix $Q$ has dimension $mn \times mn$. Indexing rows and columns by $s,t \in \{0,\dots,mn-1\}$, where index $s$ corresponds to package $\lfloor s/n \rfloor$ and layer $s \bmod n$, each entry is initialised as
\begin{equation}
Q_{s,t} = w_{\lfloor s/n\rfloor,\,\lfloor t/n\rfloor} \times c_{\,s \bmod n,\; t \bmod n}
\label{eq:qubo-init}
\end{equation}

Because a QUBO is unconstrained, the one-hot constraint~\eqref{eq:onehot} cannot be imposed directly and must instead be enforced through a penalty. We add a penalty constant $\rho$, taken as the product of the largest dependency weight, the largest cost, and the number of dependencies, so that any constraint-violating assignment is scored worse than every valid one without enumerating the assignment space. The penalised matrix is
\begin{equation}
Q' = Q + \rho E - 2\rho I_{mn}, \label{eq:penalty}
\end{equation}
where $I_{mn}$ is the $mn \times mn$ identity and $E$ is the block-diagonal matrix formed of $m$ all-ones blocks of size $n \times n$. This places the penalty exactly on the entries that couple distinct layer assignments of the same package.

The penalised QUBO is then mapped to an Ising Hamiltonian by substituting the binary values $\{0,1\}$ for the spin values $\{+1,-1\}$. The result is a sum of single-qubit Pauli-$Z$ terms and two-qubit $Z\!Z$ couplings, which defines the problem Hamiltonian $H_C$ of the QAOA ansatz.

\subsection{Impact-Guided Decomposition}
\label{sec:decomposition}

Because a package is the unit that is assigned to a layer, we decompose at the level of packages rather than individual variables: a subset of packages is optimised at a time while the layer of every other package is held fixed.

The procedure begins by assigning every package to a layer at random, which fixes all variables $x_{ik}$ and gives an initial value of the penalised objective~\eqref{eq:penalty}. Each package is then assigned an \emph{impact value}, computed by re-optimising that package's layer alone and measuring the resulting change in the objective, and the packages are sorted by impact. An \emph{active set} of packages is selected from this order and optimised with QAOA, while every package outside the active set keeps its current layer. Fixing the outside variables $x_{jl}=x_{jl}^{\ast}$ reduces the penalised objective to a sub-QUBO over the active packages,
\begin{equation}
f_{\mathcal{A}}(x)
= \sum_{s,t \in \mathcal{A}} q'_{s,t}\, x_s x_t
+ \sum_{s \in \mathcal{A}}
   \Bigg( \sum_{t \notin \mathcal{A}} (q'_{s,t} + q'_{t,s})\, x_t^{\ast} \Bigg) x_s
+ \text{const},
\label{eq:subqubo}
\end{equation}
where $\mathcal{A}$ is the set of variables of the active packages. This sub-QUBO is mapped to an Ising Hamiltonian and solved with QAOA exactly as in Sections~\ref{sec:qubo} and~\ref{sec:approach}, and the recovered layers replace the previous assignment of the active packages. The active set is advanced through the ordered packages until all have been re-optimised, after which the impacts are re-computed and the packages re-ordered. The process repeats until the objective no longer improves. Because each circuit spans only the active packages, its width is bounded independently of the system size $mn$.

An active set of $a$ packages therefore yields a sub-QUBO of $a \cdot n$ qubits, and the sub-problem size is set to keep this width within the available quantum resources.

\section{QAOA Approach for LAR}
\label{sec:approach}

\subsection{Circuit Construction and Optimisation}
\label{sec:circuit}

The Ising Hamiltonian $U_C(\gamma)$ found in Section~\ref{sec:qubo} defines the first repetition of the ansatz. We append a standard mixing Hamiltonian $U_M(\beta)$, whose role is to drive exploration of the solution space, and repeat the cost/mixer pair $p$ times to form a depth-$p$ circuit with $2p$ rotation angles. The number of qubits equals the QUBO dimension: $mn$ for the full problem, and $a \cdot n$ for an active set of $a$ packages when the impact-guided decomposition of Section~\ref{sec:decomposition} is applied.

Starting from random initial angles, we use COBYLA~\cite{powell1994}, a gradient-free optimiser, to minimise the energy expectation of the circuit and obtain the parameters $(\bm{\gamma}^\ast,\bm{\beta}^\ast)$. We then bind these parameters to the circuit, add measurements, and execute it for a fixed number of shots. The result is a distribution over bitstrings, each bitstring being a candidate package-to-layer assignment together with the number of shots in which it appeared. Under the decomposition, this construction and optimisation step is applied to each sub-QUBO in turn, with the bitstrings decoding the layers of the active packages only.

\subsection{Classical Post-Processing}
\label{sec:postproc}

Post-processing is an important step with two goals: improving the raw quantum output and formatting the result for the LAR problem. For the first goal we adopt the post-processing of the Qiskit Algorithms library~\cite{javadiabhari2024}: bitstrings whose sampling probability falls below a threshold (we use $1 \times 10^{-6}$) are discarded, and the remaining candidates are evaluated classically against the QUBO objective. As reported in our prior work~\cite{cardinal2025}, this step improves results substantially, because it recovers the best solution among the high probability states rather than relying on the single most-sampled bitstring. For the second goal, the best post-processed solution is decoded. For the full problem, the cost of the corresponding QSAP solution is recovered as
\begin{equation}
f_{\text{QSAP}} = f_q + m\rho, \label{eq:cost-recover}
\end{equation}
where $f_q$ is the QUBO cost of the chosen solution, $m$ the number of packages, and $\rho$ the penalty constant. For a valid one-hot assignment the penalty terms of~\eqref{eq:penalty} contribute a constant offset of $-m\rho$, and the $m\rho$ term removes it. The bitstring is reshaped into an $m \times n$ assignment matrix (packages as rows, layers as columns) and described in textual form as a recovered layering. Because a QUBO is unconstrained, we finally validate the decoded solution against the LAR constraints, in particular that each package is assigned to exactly one layer; solutions that assign a package to zero or more than one layer are flagged as invalid.

\section{Evaluation}
\label{sec:eval}

\subsection{Objectives and Research Questions}
\label{sec:rqs}

The goal of the evaluation is to assess and compare the performance of QAOA for recovering layered software architectures. We address four research questions, the first three concerning the monolithic QAOA pipeline and the fourth concerning its decomposition.

\begin{description}[leftmargin=2.6em,itemsep=4pt]
  \item[RQ1.] \emph{How do the algorithmic parameters of QAOA (circuit repetitions~$p$ and number of shots) affect the quality and cost of architecture recovery?} The depth $p$ and the sampling budget are the two parameters a user must set for any QAOA run. Depth controls how expressive, and how expensive, the circuit is, while the number of shots controls how many candidate solutions are sampled for classical post-processing. RQ1 measures how each affects both recovery quality (optimal rate and objective) and cost (runtime and optimiser iterations), so as to identify efficient configurations.

  \item[RQ2.] \emph{How does QAOA compare with a classical Tabu Search optimiser in terms of effectiveness and computational effort?} Tabu Search is the classical method Belle et al.~\cite{belle2015} used for this formulation, and it is our reference point. RQ2 asks whether QAOA, at the scale reachable on current simulators, matches the classical optimum and at what cost in time, situating the quantum approach relative to the established classical baseline rather than claiming a quantum advantage.

  \item[RQ3.] \emph{How robust is QAOA to quantum noise when executed on ideal versus noisy simulators?} Near-term quantum hardware is noisy, so results on an ideal simulator may not carry over to a real device. RQ3 re-runs the pipeline under simulated device noise and compares it with the ideal case, to assess how far solution quality is preserved and what noise costs in simulation effort.

  \item[RQ4.] \emph{How does impact-guided decomposition affect recovery quality and feasibility on an instance beyond the single-circuit qubit budget?} RQ1--RQ3 use a single, monolithic circuit, which does not scale to the largest instance. RQ4 asks whether solving the problem as a sequence of smaller sub-problems restores usable recovery on that instance, and how the sub-problem size and the same parameters studied in RQ1 behave under decomposition.
\end{description}

\subsection{Experimental Setup}
\label{sec:setup}

\paragraph{Subject systems.}
We evaluate the approach on three real-world Java systems. The monolithic QAOA experiments (RQ1--RQ3) recover all three systems with $n=3$ layers:
\begin{itemize}[leftmargin=1.4em,itemsep=1pt]
  \item \emph{JReversePro 1.0}: 5 packages, 723 dependencies ($15$ qubits);
  \item \emph{JReversePro 1.2}: 6 packages, 702 dependencies ($18$ qubits);
  \item \emph{JUnit 4.13.2}: 8 packages, 1538 dependencies ($24$ qubits).
        
\end{itemize}
The number of qubits equals the QUBO dimension $m \cdot n$, so the three systems span $15$, $18$, and $24$ qubits respectively.

\paragraph{Platforms.}
Quantum experiments were executed on the Qiskit AER simulator. For RQ3 we additionally used Qiskit AER with the IBM Quebec noise model, as well as depolarising and dephasing noise models.

The classical baseline is the Tabu Search QSAP solver used by Belle et al.~\cite{belle2015}, which we adopt for direct comparability with the original formulation. Each Tabu Search run uses a tabu-list length of $10$, except for JReversePro 1.0 where a length of $8$ was used to avoid the algorithm stalling; Tabu Search was run for $1000$ iterations.

\paragraph{Compute environment.}
All runs were executed on the Narval cluster of Calcul Qu\'ebec, using all cores of a compute node. Each node provides two AMD EPYC 7532 CPUs ($64$ cores / $128$ threads) and approximately $200$~GB of RAM.

\paragraph{Constant parameters.}
The LAR cost parameters were fixed at $\alpha=1$ (skip), $\beta=15$ (back), and $\gamma=2$ (intra).

Circuit transpilation used the heaviest optimisation level with $1^{\circ}$ angle approximation; circuit parameters were initialised randomly. COBYLA used its default tolerance and a maximum of $1000$ iterations. The post-processing sampling-probability threshold was $1 \times 10^{-6}$.

\subsection{Experiment Design and Metrics}
\label{sec:design}

To account for the stochasticity of the classical optimiser and of circuit measurement, every experiment is repeated $30$ times. Each run (i) generates a QUBO model from the dependency graph, (ii) constructs the quantum circuit, (iii) solves it with QAOA and improves the solution by post-processing, and (iv) decodes the result into a layer assignment. We collect the following metrics, averaged over the $30$ runs:
\begin{itemize}[leftmargin=1.4em,itemsep=1pt]
  \item \emph{Avg.\ Time (s)}: runtime per run, from QUBO creation through
        post-processing, excluding server waiting time.
  \item \emph{Iterations}: number of cost-function evaluations performed by the
        classical optimiser.
  \item \emph{States Evaluated}: number of distinct candidate states evaluated
        classically in post-processing; the percentage in parentheses is the
        fraction of the full $2^{mn}$ assignment space that this represents.
  \item \emph{Avg.\ Objective}: average over runs of the minimum layering cost
        found in each run.
  \item \emph{Optimal Rate}: the percentage of the $30$ runs that reached the
        best assignment known for that instance.
        
\end{itemize}

\section{Results}
\label{sec:results}

\subsection{RQ1: Parameter Sensitivity}
\label{sec:rq1}

For RQ1 we vary the number of circuit repetitions $p$ (from $1$ to $10$) and the number of shots on the ideal AER simulator, for the monolithic (non-decomposed) QAOA pipeline. Tables~\ref{tab:jrp10}, \ref{tab:jrp12}, and~\ref{tab:junit24} report the results for the three systems.

Three patterns emerge consistently. First, \emph{the number of repetitions has a large effect on runtime but little, and inconsistent, effect on accuracy.} Increasing $p$ raises execution time sharply---on JReversePro~1.0 at $250{,}000$ shots the average time rises from $1.30$~s at $p{=}1$ to $12.34$~s at $p{=}10$---yet the optimal rate does not improve monotonically and sometimes falls. Deeper circuits do not ensure better solutions in this regime.

Second, \emph{the number of shots is the dominant lever on solution quality.} Higher shot counts increase the optimal rate at moderate additional time cost. The effect is clearest on JReversePro~1.2 (Table~\ref{tab:jrp12}): at $150{,}000$ shots the optimal rate stays between $53\%$ and $73\%$, whereas at $1{,}000{,}000$ shots it reaches $97$--$100\%$, with two configurations ($p{=}7$ and $p{=}10$) recovering the optimum in every run.

Third, \emph{solution quality tracks the fraction of the assignment space covered by classical post-processing.} The ``States Evaluated'' percentage is the share of the $2^{mn}$ space evaluated after thresholding, and it correlates strongly with the optimal rate: low coverage accompanies poor results. Higher shot counts produce more distinct high-probability states for post-processing to evaluate, which is the mechanism by which shots improve accuracy. This reliance is most visible at scale: on the $24$-qubit JUnit~4.13.2 instance (Table~\ref{tab:junit24}), even $1{,}000{,}000$ shots cover only about $5.6\%$ of the assignment space and the optimal rate never exceeds $6.7\%$. Raising the shot count to one million therefore did not lift JUnit out of low-accuracy territory. We did not exceed one million shots, so as to keep the setup representative of what would be feasible on real quantum hardware.

\begin{table}[!htbp]
\centering
\caption{Impact of circuit repetitions and shots on QAOA recovery
of JReversePro 1.0 (15 qubits), using the Qiskit AER ideal
simulator (30 runs per configuration).}
\label{tab:jrp10}
\footnotesize
\setlength{\tabcolsep}{3pt}
\begin{tabular}{@{}rrccccc@{}}
\toprule
Shots & Rep. & Time (s) & Iter. & States eval. & Avg. obj. & Opt. (\%) \\
\midrule
\multirow{6}{*}{150k}
 & 1  & 1.14 $\pm$ 0.05  & 30 $\pm$ 2  & 26101 (80\%) & 570 $\pm$ 22 & 87 \\
 & 2  & 1.58 $\pm$ 0.09  & 39 $\pm$ 3  & 25974 (79\%) & 570 $\pm$ 22 & 87 \\
 & 3  & 2.35 $\pm$ 0.16  & 54 $\pm$ 3  & 26306 (80\%) & 572 $\pm$ 26 & 83 \\
 & 5  & 4.30 $\pm$ 0.36  & 83 $\pm$ 7  & 26702 (81\%) & 572 $\pm$ 24 & 83 \\
 & 7  & 6.79 $\pm$ 0.37  & 109 $\pm$ 5 & 26801 (82\%) & 570 $\pm$ 25 & 87 \\
 & 10 & 20.56 $\pm$ 1.14 & 151 $\pm$ 7 & 26879 (82\%) & 578 $\pm$ 29 & 73 \\
\midrule
\multirow{6}{*}{200k}
 & 1  & 1.21 $\pm$ 0.06  & 29 $\pm$ 2  & 26944 (82\%) & 574 $\pm$ 26 & 80 \\
 & 2  & 1.65 $\pm$ 0.09  & 39 $\pm$ 3  & 27560 (84\%) & 578 $\pm$ 29 & 73 \\
 & 3  & 2.37 $\pm$ 0.16  & 53 $\pm$ 4  & 27531 (84\%) & 569 $\pm$ 22 & 87 \\
 & 5  & 4.34 $\pm$ 0.24  & 82 $\pm$ 4  & 27919 (85\%) & 567 $\pm$ 20 & 90 \\
 & 7  & 6.98 $\pm$ 0.41  & 111 $\pm$ 7 & 28084 (86\%) & 570 $\pm$ 22 & 87 \\
 & 10 & 11.97 $\pm$ 0.53 & 151 $\pm$ 7 & 28130 (86\%) & 574 $\pm$ 26 & 80 \\
\midrule
\multirow{6}{*}{250k}
 & 1  & 1.30 $\pm$ 0.07  & 30 $\pm$ 3  & 28347 (87\%) & 572 $\pm$ 24 & 83 \\
 & 2  & 1.72 $\pm$ 0.11  & 39 $\pm$ 4  & 27974 (85\%) & 570 $\pm$ 22 & 87 \\
 & 3  & 2.43 $\pm$ 0.10  & 53 $\pm$ 3  & 28167 (86\%) & 570 $\pm$ 22 & 87 \\
 & 5  & 4.49 $\pm$ 0.37  & 82 $\pm$ 5  & 28795 (88\%) & 567 $\pm$ 20 & 90 \\
 & 7  & 7.05 $\pm$ 0.43  & 110 $\pm$ 6 & 28909 (88\%) & 565 $\pm$ 16 & 93 \\
 & 10 & 12.34 $\pm$ 0.67 & 152 $\pm$ 8 & 28958 (88\%) & 567 $\pm$ 20 & 90 \\
\bottomrule
\end{tabular}
\end{table}

\begin{table}[!htbp]
\centering
\caption{Impact of circuit repetitions and shots on QAOA recovery
of JReversePro 1.2 (18 qubits), using the Qiskit AER ideal
simulator (30 runs per configuration).}
\label{tab:jrp12}
\footnotesize
\setlength{\tabcolsep}{3pt}
\begin{tabular}{@{}rrccccc@{}}
\toprule
Shots & Rep. & Time (s) & Iter. & States eval. & Avg. obj. & Opt. (\%) \\
\midrule
\multirow{6}{*}{150k}
 & 1  & 3.39 $\pm$ 0.15  & 30 $\pm$ 3   & 90138 (34\%)  & 318 $\pm$ 26 & 53 \\
 & 2  & 4.46 $\pm$ 0.23  & 38 $\pm$ 3   & 90315 (34\%)  & 316 $\pm$ 27 & 73 \\
 & 3  & 3.95 $\pm$ 0.26  & 53 $\pm$ 4   & 91307 (35\%)  & 320 $\pm$ 36 & 63 \\
 & 5  & 7.18 $\pm$ 0.44  & 84 $\pm$ 6   & 92918 (35\%)  & 315 $\pm$ 34 & 70 \\
 & 7  & 11.25 $\pm$ 0.74 & 111 $\pm$ 7  & 94828 (36\%)  & 318 $\pm$ 32 & 67 \\
 & 10 & 19.20 $\pm$ 0.78 & 150 $\pm$ 6  & 95397 (36\%)  & 324 $\pm$ 60 & 60 \\
\midrule
\multirow{6}{*}{200k}
 & 1  & 2.26 $\pm$ 0.10  & 30 $\pm$ 3   & 105198 (40\%) & 321 $\pm$ 58 & 80 \\
 & 2  & 3.06 $\pm$ 0.26  & 40 $\pm$ 4   & 104864 (40\%) & 306 $\pm$ 3  & 90 \\
 & 3  & 4.20 $\pm$ 0.20  & 54 $\pm$ 3   & 108949 (42\%) & 314 $\pm$ 17 & 67 \\
 & 5  & 7.30 $\pm$ 0.41  & 82 $\pm$ 6   & 111294 (42\%) & 323 $\pm$ 44 & 63 \\
 & 7  & 11.44 $\pm$ 0.68 & 111 $\pm$ 7  & 111456 (43\%) & 317 $\pm$ 48 & 77 \\
 & 10 & 33.69 $\pm$ 1.72 & 151 $\pm$ 7  & 113431 (43\%) & 306 $\pm$ 8  & 97 \\
\midrule
\multirow{6}{*}{250k}
 & 1  & 4.01 $\pm$ 0.20  & 29 $\pm$ 2   & 124072 (47\%) & 307 $\pm$ 4  & 77 \\
 & 2  & 3.17 $\pm$ 0.14  & 38 $\pm$ 2   & 122388 (47\%) & 312 $\pm$ 25 & 83 \\
 & 3  & 4.38 $\pm$ 0.25  & 53 $\pm$ 3   & 121831 (46\%) & 312 $\pm$ 13 & 63 \\
 & 5  & 13.05 $\pm$ 0.97 & 83 $\pm$ 6   & 126016 (48\%) & 307 $\pm$ 3  & 83 \\
 & 7  & 11.77 $\pm$ 0.66 & 112 $\pm$ 7  & 126636 (48\%) & 322 $\pm$ 58 & 83 \\
 & 10 & 33.18 $\pm$ 1.80 & 150 $\pm$ 7  & 127978 (49\%) & 311 $\pm$ 12 & 67 \\
\midrule
\multirow{6}{*}{500k}
 & 1  & 5.19 $\pm$ 0.34  & 28 $\pm$ 2   & 166205 (63\%) & 308 $\pm$ 8  & 80 \\
 & 2  & 6.50 $\pm$ 0.35  & 39 $\pm$ 4   & 165034 (63\%) & 306 $\pm$ 3  & 83 \\
 & 3  & 5.01 $\pm$ 0.20  & 52 $\pm$ 2   & 164020 (63\%) & 307 $\pm$ 8  & 90 \\
 & 5  & 8.71 $\pm$ 1.08  & 84 $\pm$ 7   & 168563 (64\%) & 307 $\pm$ 8  & 93 \\
 & 7  & 12.49 $\pm$ 0.82 & 112 $\pm$ 9  & 171463 (65\%) & 307 $\pm$ 4  & 80 \\
 & 10 & 37.95 $\pm$ 2.43 & 151 $\pm$ 10 & 171896 (66\%) & 306 $\pm$ 3  & 90 \\
\midrule
\multirow{6}{*}{1M}
 & 1  & 4.15 $\pm$ 0.22  & 29 $\pm$ 3   & 198195 (76\%) & 305 $\pm$ 2  & 97 \\
 & 2  & 4.91 $\pm$ 0.33  & 38 $\pm$ 4   & 201371 (77\%) & 305 $\pm$ 2  & 97 \\
 & 3  & 6.04 $\pm$ 0.27  & 52 $\pm$ 3   & 201786 (77\%) & 306 $\pm$ 2  & 93 \\
 & 5  & 9.49 $\pm$ 0.58  & 83 $\pm$ 6   & 206435 (79\%) & 305 $\pm$ 2  & 97 \\
 & 7  & 23.32 $\pm$ 1.02 & 111 $\pm$ 6  & 206898 (79\%) & 305 $\pm$ 0  & 100 \\
 & 10 & 39.05 $\pm$ 3.90 & 153 $\pm$ 8  & 207590 (79\%) & 305 $\pm$ 0  & 100 \\
\bottomrule
\end{tabular}
\end{table}


\begin{table}[!htbp]
\centering
\caption{Impact of circuit repetitions and shots on QAOA recovery
of JUnit 4.13.2 (24 qubits), using the Qiskit AER ideal
simulator (30 runs per configuration).}
\label{tab:junit24}
\footnotesize
\setlength{\tabcolsep}{3pt}
\begin{tabular}{@{}rrccccc@{}}
\toprule
Shots & Rep. & Time (s) & Iter. & States eval. & Avg. obj. & Opt. (\%) \\
\midrule
\multirow{6}{*}{250k}
 & 1  & 22.61 $\pm$ 1.41   & 29 $\pm$ 2   & 243883 (1.5\%) & 2137 $\pm$ 181 & 0   \\
 & 2  & 26.41 $\pm$ 2.29   & 39 $\pm$ 4   & 244021 (1.5\%) & 2195 $\pm$ 228 & 3.3 \\
 & 3  & 44.58 $\pm$ 2.49   & 53 $\pm$ 3   & 244592 (1.5\%) & 2121 $\pm$ 184 & 3.3 \\
 & 5  & 153.24 $\pm$ 6.41  & 82 $\pm$ 3   & 245638 (1.5\%) & 2151 $\pm$ 233 & 0   \\
 & 7  & 169.93 $\pm$ 12.54 & 112 $\pm$ 9  & 246042 (1.5\%) & 2173 $\pm$ 164 & 0   \\
 & 10 & 526.44 $\pm$ 59.49 & 160 $\pm$ 16 & 246186 (1.5\%) & 2146 $\pm$ 160 & 0   \\
\midrule
\multirow{6}{*}{1M}
 & 1  & 23.41 $\pm$ 1.07   & 28 $\pm$ 2   & 907857 (5.4\%) & 1954 $\pm$ 109 & 3.3 \\
 & 2  & 54.67 $\pm$ 2.39   & 38 $\pm$ 3   & 908855 (5.4\%) & 1936 $\pm$ 89  & 3.3 \\
 & 3  & 52.50 $\pm$ 2.57   & 53 $\pm$ 3   & 920893 (5.5\%) & 1969 $\pm$ 110 & 3.3 \\
 & 5  & 106.02 $\pm$ 6.50  & 83 $\pm$ 6   & 926446 (5.5\%) & 1991 $\pm$ 137 & 6.7 \\
 & 7  & 176.95 $\pm$ 10.38 & 111 $\pm$ 7  & 939001 (5.6\%) & 1990 $\pm$ 127 & 6.7 \\
 & 10 & 545.17 $\pm$ 45.61 & 158 $\pm$ 13 & 941779 (5.6\%) & 1988 $\pm$ 133 & 3.3 \\
\bottomrule
\end{tabular}
\end{table}


\noindent\emph{Summary (RQ1).} The number of shots, not circuit depth, governs solution quality, and its effect operates through the coverage achieved by classical post-processing. A shallow circuit ($p{=}1$ or $2$) with a shot budget scaled to the problem size is the most efficient and stable configuration. The approach is accurate on the $15$- and $18$-qubit systems but does not reach usable accuracy on the $24$-qubit system within a hardware-realistic shot budget.

\subsection{RQ2: Comparison with a Classical Optimiser}
\label{sec:rq2}

Table~\ref{tab:compare} compares the best QAOA configuration from RQ1 against Tabu Search on each system. As expected, QAOA does not match the accuracy of Tabu Search overall. It reaches the same optimal rate only on JReversePro 1.2, where there are two distinct assignments that achieve the optimum, giving QAOA two routes to a correct answer. On the larger JUnit instance the gap is wide: QAOA reaches the optimum in only $6.7\%$ of runs against $100\%$ for Tabu Search, and its average objective ($1990$) is above the optimum ($1867$). In runtime the gap is equally clear and unsurprising: Tabu Search completes in well under a second, whereas QAOA takes seconds on the small systems and close to three minutes per run on JUnit. We note that we did not tune the COBYLA termination tolerance, which could affect both runtime and accuracy.

\begin{table}[t]
\centering
\small
\caption{QAOA versus Tabu Search on cost, runtime, and optimiser effort,
30 executions per configuration. QAOA uses the best RQ1 configuration on the
ideal AER simulator.}\label{tab:compare}
\begin{tabular}{@{}l l r r r r@{}}
\toprule
Program & Approach & Avg.\ Time (s) & Iterations & Avg.\ Obj. & Opt.\ Rate\\
\midrule
JReversePro 1.0 & Tabu Search & 0.07 $\pm$ 0.00 & 1000 & 561 $\pm$ 0  & 100\%\\
JReversePro 1.0 & QAOA        & 7.05 $\pm$ 0.43 & 110 $\pm$ 6 & 565 $\pm$ 16 & 93\%\\
JReversePro 1.2 & Tabu Search & 0.09 $\pm$ 0.00 & 1000 & 305 $\pm$ 0  & 100\%\\
JReversePro 1.2 & QAOA        & 23.32 $\pm$ 1.02 & 111 $\pm$ 6 & 305 $\pm$ 0  & 100\%\\
JUnit 4.13.2    & Tabu Search & 0.11 $\pm$ 0.00 & 1000 & 1867 $\pm$ 0 & 100\%\\
JUnit 4.13.2    & QAOA        & 176.95 $\pm$ 10.38 & 111 $\pm$ 7 & 1990 $\pm$ 127 & 6.7\%\\
\bottomrule
\end{tabular}
\end{table}

\noindent\emph{Summary (RQ2).} At the scale reachable on current simulators, QAOA is competitive with the classical heuristic only on the smallest and most degenerate instances and is uniformly slower. We observe no quantum advantage, which is the expected outcome in the NISQ regime. Thus the contribution of the quantum formulation lies in establishing a working pipeline for the problem rather than in outperforming classical search today.

\subsection{RQ3: Robustness Across Execution Environments}
\label{sec:rq3}

For RQ3 we re-ran the monolithic QAOA pipeline under simulated noise and compared it to the ideal simulator. Table~\ref{tab:noise} reports the results.

On the $15$-qubit JReversePro 1.0, QAOA was robust in solution quality under all three noise models: it recovered the optimal layering ($561 \pm 0$) in every run, slightly exceeding the $93\%$ optimal rate of the ideal simulator. The cost of noise is paid almost entirely in simulation time, which rose by roughly two orders of magnitude from $7.05$~s on the ideal simulator to $490$--$921$~s under the various noise models, reflecting the much greater expense of simulating noisy dynamics. For this small instance, the dominant practical obstacle is therefore simulation cost, not solution degradation.

For the $18$-qubit (JReversePro 1.2) and $24$-qubit (JUnit 4.13.2) instances, noisy simulation did not complete within the available compute budget. These cases mark the practical ceiling of noisy simulation in our setup: beyond about fifteen qubits, the combination of noise simulation and the high shot counts required for accuracy became infeasible on a single Narval node. This boundary, rather than any single accuracy figure, is the central finding of RQ3 and directly motivates the error-mitigation and real-hardware directions discussed below. This infeasibility is specific to the monolithic pipeline; RQ4 shows that decomposition restores noisy execution on the same instance.

\begin{table}[t]
\centering
\small
\caption{Effect of simulated noise on cost, runtime, and optimiser effort,
30 executions per configuration. ``---'' denotes runs that did not complete
within the available compute budget.}\label{tab:noise}
\begin{tabular}{@{}l l r r r r@{}}
\toprule
Program & Simulator & Avg.\ Time (s) & Iterations & Avg.\ Obj. & Opt.\ Rate\\
\midrule
JReversePro 1.0 & Ideal              & 7.05 $\pm$ 0.43   & 110 $\pm$ 6 & 565 $\pm$ 16 & 93\%\\
JReversePro 1.0 & IBM Quebec noise   & 921.03 $\pm$ 42.87 & 112 $\pm$ 7 & 561 $\pm$ 0  & 100\%\\
JReversePro 1.0 & Depolarising noise & 490.34 $\pm$ 14.87 & 115 $\pm$ 8 & 561 $\pm$ 0  & 100\%\\
JReversePro 1.0 & Dephasing noise    & 769.88 $\pm$ 37.37 & 112 $\pm$ 8 & 561 $\pm$ 0  & 100\%\\
JReversePro 1.2 & Ideal              & 23.32 $\pm$ 1.02  & 111 $\pm$ 6 & 305 $\pm$ 0  & 100\%\\
JReversePro 1.2 & IBM Quebec noise   & ---  & --- & --- & ---\\
JUnit 4.13.2    & Ideal              & 176.95 $\pm$ 10.38 & 111 $\pm$ 7 & 1990 $\pm$ 127 & 6.7\%\\
JUnit 4.13.2    & IBM Quebec noise   & ---  & --- & --- & ---\\
\bottomrule
\end{tabular}
\end{table}

\subsection{RQ4: Impact-Guided Decomposition}
\label{sec:rq4}

RQ1 established that the monolithic pipeline fails on the $24$-qubit JUnit instance, reaching the optimum in at most $6.7\%$ of runs (Table~\ref{tab:junit24}). For RQ4 we apply the impact-guided decomposition of Section~\ref{sec:decomposition} to that same instance, so that each QAOA circuit spans only an active set of packages. We report results for four active-set configurations, on the ideal simulator and under the IBM Montreal noise model (Tables~\ref{tab:igdec-6-ideal}--\ref{tab:igdec-12-montreal}). An active set of $a$ packages over $n$ layers yields a sub-QUBO of $a\cdot n$ qubits, so the two dimensions co-vary across these tables: the $6$- and $9$-qubit configurations use $n=3$ layers (three weight matrices, $a=2$ and $a=3$), and the $8$- and $12$-qubit configurations use $n=4$ layers (four weight matrices, $a=2$ and $a=3$). Because the best known recovery cost is the same, $1867$, at both three and four layers, the optimal rates of the decomposed and monolithic runs are measured against a common target and are directly comparable.

\paragraph{Accuracy depends on the sub-problem size, not the number of repetitions.}
Decomposition changes the picture substantially. The best optimal rate rises monotonically with the number of qubits in the sub-problem: $30\%$ at $6$ qubits (Table~\ref{tab:igdec-6-ideal}), $70\%$ at $8$ qubits (Table~\ref{tab:igdec-8-ideal}), $83.3\%$ at $9$ qubits (Table~\ref{tab:igdec-9-ideal}), and $93.3\%$ at $12$ qubits (Table~\ref{tab:igdec-12-ideal}). The decisive comparison is that the $9$-qubit, three-layer configuration ($83.3\%$) outperforms the $8$-qubit, four-layer one ($70\%$):  recovery quality depends on the total size of the active sub-problem, not the number of layers as such. A larger active set helps because it moves more packages jointly at each step of the greedy decomposition, and additional layers help, at a fixed number of packages, by enriching the set of placements each package can take; both enlarge the sub-problem. The best configuration, $12$ qubits ($a=3$, $n=4$), is the one that maximises both. There, at $2{,}000$ shots and $p{=}1$, QAOA recovers the optimal layering in $93.3\%$ of runs with an average cost of $1867\pm1$, against $6.7\%$ and $1990$ for the monolithic pipeline on the same instance (Table~\ref{tab:compare}). Smaller active sets reach the optimum only intermittently and with far greater dispersion (for example $1918\pm129$ at best for $9$ qubits), and the $6$-qubit configuration, which frees only two of the eight packages at a time, largely stalls at a cost of $2091$--$2122$.

\paragraph{Solution quality decreases with the shot budget.}
The most striking effect is that increasing the number of shots \emph{degrades} the optimal rate, reversing the dominant trend of RQ1. At $12$ qubits and $p{=}1$, the optimal rate falls from $93.3\%$ at $2{,}000$ shots to $80.0\%$ at $10{,}000$ and $66.7\%$ at $200{,}000$; the same monotone decline holds for every active-set size and in the noisy runs. The number of COBYLA iterations is essentially constant across shot budgets (about $2{,}630$ for the $12$-qubit, $p{=}1$ runs), so the shot count changes only the precision with which each sub-problem's expectation is estimated. A plausible explanation follows from the structure of the method. Impact-guided decomposition optimises only a few packages at a time while holding the rest fixed, so each step is greedy, each sub-problem is then small enough to solve almost exactly, and the quantum circuit is not the bottleneck. With a large shot budget, every step commits to the locally optimal move and the procedure is readily trapped in a poor global layering; with a small budget, sampling noise perturbs these moves and lets the sweep escape such local optima. Consistently with this reading, the ``states evaluated'' column rises with the shot count while the optimal rate falls, so the classical-coverage mechanism that explained RQ1 does not account for the behaviour under decomposition. Practically, decomposition attains its best accuracy with the \emph{smallest} shot budget tested, which makes it inexpensive in both circuit depth and sampling.

\paragraph{Circuit depth has no consistent effect.}
As in RQ1, and consistently with prior reports on IGDec-QAOA, the circuit depth $p$ does not systematically improve recovery: within each active-set size the optimal rate is flat or slightly decreasing in $p$ while runtime grows with it. At $12$ qubits and $2{,}000$ shots, $p{=}1$ is simultaneously the cheapest, among the most accurate, and the most stable configuration, recovering the optimum at $1867\pm1$, against $1880\pm68$ at $p{=}10$. We therefore retain $p{=}1$ as the reference depth throughout.

\paragraph{Behaviour under noise.}
Under the IBM Montreal noise model, the recommended $12$-qubit configuration keeps its quality: at $2{,}000$ shots and $p{=}1$ it reaches $96.7\%$ ($1879\pm68$), slightly above the ideal-simulator value. The smaller configurations lose some ground under noise, so both robustness and accuracy improve with active-set size. The main cost of noise is time: a run at $12$ qubits rises from about $70$~s to about $11{,}600$~s, more than a hundredfold. This is also why the $12$-qubit noisy runs cover only the lowest depths and shot budgets; the larger settings were too expensive to run on a single node.

\noindent\emph{Summary (RQ4).} Impact-guided decomposition is what makes QAOA work on the instance where the monolithic pipeline fails: it restores reliable recovery (up to $93.3\%$ on the ideal simulator and $96.7\%$ under noise, against $6.7\%$ for the monolithic pipeline) and keeps noisy execution feasible on the $24$-qubit system,  by limiting each circuit to a small active set. Recovery quality depends on the size of the active sub-problem rather than the number of layers, improves as the shot budget is \emph{reduced}, and is insensitive to circuit depth. It is the way decomposition restructures the search, not any increase in sampling, that carries QAOA past the qubit ceiling of RQ1 and RQ3.

\begin{table}[!htbp]
\centering
\caption{IGDec-QAOA recovery of JUnit 4.13.2 with an active set of
6 qubits (3 weight matrices), using the Qiskit AER ideal simulator
(30 runs per configuration).}
\label{tab:igdec-6-ideal}
\footnotesize
\setlength{\tabcolsep}{3pt}
\begin{tabular}{@{}rrccccc@{}}
\toprule
Shots & Rep. & Time (s) & Iter. & States eval. & Avg. obj. & Opt. (\%) \\
\midrule
\multirow{6}{*}{2k}
 & 1  & 46.97 $\pm$ 4.90   & 3496 $\pm$ 26  & 6996 (0.042\%) & 2091 $\pm$ 165 & 30.0 \\
 & 2  & 96.31 $\pm$ 33.86  & 4657 $\pm$ 36  & 6963 (0.042\%) & 2091 $\pm$ 172 & 30.0 \\
 & 3  & 100.70 $\pm$ 13.44 & 6305 $\pm$ 39  & 7037 (0.042\%) & 2110 $\pm$ 156 & 23.3 \\
 & 5  & 181.08 $\pm$ 7.94  & 9670 $\pm$ 54  & 7106 (0.042\%) & 2110 $\pm$ 156 & 23.3 \\
 & 7  & 291.20 $\pm$ 14.03 & 13056 $\pm$ 41 & 7124 (0.042\%) & 2122 $\pm$ 151 & 20.0 \\
 & 10 & 506.87 $\pm$ 28.20 & 18154 $\pm$ 76 & 7122 (0.042\%) & 2122 $\pm$ 142 & 16.7 \\
\midrule
\multirow{6}{*}{10k}
 & 1  & 47.04 $\pm$ 0.91   & 3483 $\pm$ 22  & 7437 (0.044\%) & 2122 $\pm$ 142 & 16.7 \\
 & 2  & 68.51 $\pm$ 2.71   & 4657 $\pm$ 38  & 7462 (0.044\%) & 2122 $\pm$ 142 & 16.7 \\
 & 3  & 99.77 $\pm$ 3.50   & 6302 $\pm$ 31  & 7483 (0.045\%) & 2110 $\pm$ 148 & 20.0 \\
 & 5  & 179.74 $\pm$ 7.04  & 9673 $\pm$ 41  & 7509 (0.045\%) & 2129 $\pm$ 143 & 16.7 \\
 & 7  & 309.11 $\pm$ 11.98 & 13038 $\pm$ 49 & 7517 (0.045\%) & 2123 $\pm$ 142 & 16.7 \\
 & 10 & 518.70 $\pm$ 22.54 & 18144 $\pm$ 58 & 7521 (0.045\%) & 2122 $\pm$ 142 & 16.7 \\
\midrule
\multirow{6}{*}{200k}
 & 1  & 132.24 $\pm$ 11.54 & 3494 $\pm$ 25  & 7644 (0.046\%) & 2122 $\pm$ 142 & 16.7 \\
 & 2  & 131.35 $\pm$ 3.43  & 4650 $\pm$ 34  & 7656 (0.046\%) & 2122 $\pm$ 142 & 16.7 \\
 & 3  & 175.61 $\pm$ 5.05  & 6292 $\pm$ 36  & 7655 (0.046\%) & 2122 $\pm$ 142 & 16.7 \\
 & 5  & 250.09 $\pm$ 9.82  & 9683 $\pm$ 45  & 7667 (0.046\%) & 2122 $\pm$ 142 & 16.7 \\
 & 7  & 346.89 $\pm$ 6.32  & 13051 $\pm$ 47 & 7667 (0.046\%) & 2122 $\pm$ 142 & 16.7 \\
 & 10 & 572.44 $\pm$ 19.81 & 18154 $\pm$ 75 & 7667 (0.046\%) & 2122 $\pm$ 142 & 16.7 \\
\bottomrule
\end{tabular}
\end{table}

\begin{table}[!htbp]
\centering
\caption{IGDec-QAOA recovery of JUnit 4.13.2 with an active set of
6 qubits (3 weight matrices), using the Qiskit AER simulator with the
IBM Montreal noise model (30 runs per configuration).}
\label{tab:igdec-6-montreal}
\footnotesize
\setlength{\tabcolsep}{3pt}
\begin{tabular}{@{}rrccccc@{}}
\toprule
Shots & Rep. & Time (s) & Iter. & States eval. & Avg. obj. & Opt. (\%) \\
\midrule
\multirow{6}{*}{2k}
 & 1  & 1105.76 $\pm$ 47.81  & 3494 $\pm$ 25   & 7393 (0.044\%) & 2122 $\pm$ 142 & 16.7 \\
 & 2  & 1571.03 $\pm$ 46.77  & 4651 $\pm$ 39   & 7480 (0.045\%) & 2122 $\pm$ 142 & 16.7 \\
 & 3  & 2159.50 $\pm$ 105.38 & 6302 $\pm$ 25   & 7558 (0.045\%) & 2122 $\pm$ 142 & 16.7 \\
 & 5  & 3170.06 $\pm$ 49.17  & 9688 $\pm$ 46   & 7635 (0.046\%) & 2122 $\pm$ 142 & 16.7 \\
 & 7  & 4701.07 $\pm$ 143.89 & 13079 $\pm$ 58  & 7664 (0.046\%) & 2122 $\pm$ 142 & 16.7 \\
 & 10 & 6462.46 $\pm$ 73.77  & 18250 $\pm$ 96  & 7676 (0.046\%) & 2122 $\pm$ 142 & 16.7 \\
\midrule
\multirow{6}{*}{10k}
 & 1  & 1090.04 $\pm$ 17.05  & 3489 $\pm$ 29   & 7604 (0.045\%) & 2122 $\pm$ 142 & 16.7 \\
 & 2  & 1523.38 $\pm$ 57.16  & 4649 $\pm$ 27   & 7646 (0.046\%) & 2122 $\pm$ 142 & 16.7 \\
 & 3  & 2047.01 $\pm$ 90.62  & 6289 $\pm$ 34   & 7668 (0.046\%) & 2122 $\pm$ 142 & 16.7 \\
 & 5  & 3190.89 $\pm$ 77.82  & 9688 $\pm$ 48   & 7679 (0.046\%) & 2122 $\pm$ 142 & 16.7 \\
 & 7  & 4393.38 $\pm$ 103.30 & 13085 $\pm$ 69  & 7680 (0.046\%) & 2122 $\pm$ 142 & 16.7 \\
 & 10 & 6889.96 $\pm$ 197.35 & 18218 $\pm$ 101 & 7676 (0.046\%) & 2122 $\pm$ 142 & 16.7 \\
\midrule
\multirow{6}{*}{250k}
 & 1  & 1212.48 $\pm$ 21.52  & 3494 $\pm$ 32   & 7677 (0.046\%) & 2122 $\pm$ 142 & 16.7 \\
 & 2  & 1590.20 $\pm$ 21.55  & 4649 $\pm$ 33   & 7679 (0.046\%) & 2122 $\pm$ 142 & 16.7 \\
 & 3  & 2138.98 $\pm$ 58.78  & 6289 $\pm$ 33   & 7680 (0.046\%) & 2122 $\pm$ 142 & 16.7 \\
 & 5  & 3319.21 $\pm$ 142.14 & 9691 $\pm$ 41   & 7680 (0.046\%) & 2122 $\pm$ 142 & 16.7 \\
 & 7  & 4585.35 $\pm$ 159.21 & 13076 $\pm$ 73  & 7680 (0.046\%) & 2122 $\pm$ 142 & 16.7 \\
 & 10 & 6456.84 $\pm$ 134.51 & 18246 $\pm$ 93  & 7680 (0.046\%) & 2122 $\pm$ 142 & 16.7 \\
\midrule
\multirow{6}{*}{1M}
 & 1  & 1567.62 $\pm$ 27.02  & 3489 $\pm$ 29   & 7680 (0.046\%) & 2122 $\pm$ 142 & 16.7 \\
 & 2  & 1965.57 $\pm$ 19.55  & 4646 $\pm$ 26   & 7680 (0.046\%) & 2122 $\pm$ 142 & 16.7 \\
 & 3  & 2435.77 $\pm$ 64.12  & 6303 $\pm$ 38   & 7680 (0.046\%) & 2122 $\pm$ 142 & 16.7 \\
 & 5  & 3558.71 $\pm$ 58.79  & 9682 $\pm$ 43   & 7680 (0.046\%) & 2122 $\pm$ 142 & 16.7 \\
 & 7  & 4829.78 $\pm$ 236.96 & 13076 $\pm$ 67  & 7680 (0.046\%) & 2122 $\pm$ 142 & 16.7 \\
 & 10 & 6917.31 $\pm$ 248.58 & 18236 $\pm$ 86  & 7680 (0.046\%) & 2122 $\pm$ 142 & 16.7 \\
\bottomrule
\end{tabular}
\end{table}

\begin{table}[!htbp]
\centering
\caption{IGDec-QAOA recovery of JUnit 4.13.2 with an active set of
9 qubits (3 weight matrices), using the Qiskit AER ideal simulator
(30 runs per configuration).}
\label{tab:igdec-9-ideal}
\footnotesize
\setlength{\tabcolsep}{3pt}
\begin{tabular}{@{}rrccccc@{}}
\toprule
Shots & Rep. & Time (s) & Iter. & States eval. & Avg. obj. & Opt. (\%) \\
\midrule
\multirow{6}{*}{2k}
 & 1  & 45.02 $\pm$ 1.85   & 2625 $\pm$ 20  & 21648 (0.13\%) & 1942 $\pm$ 152 & 76.7 \\
 & 2  & 61.68 $\pm$ 2.08   & 3481 $\pm$ 27  & 20884 (0.12\%) & 1942 $\pm$ 152 & 80.0 \\
 & 3  & 87.22 $\pm$ 1.42   & 4717 $\pm$ 30  & 21063 (0.13\%) & 1942 $\pm$ 152 & 80.0 \\
 & 5  & 200.80 $\pm$ 8.37  & 7271 $\pm$ 46  & 21687 (0.13\%) & 1942 $\pm$ 152 & 80.0 \\
 & 7  & 290.45 $\pm$ 6.82  & 9799 $\pm$ 60  & 22168 (0.13\%) & 1918 $\pm$ 129 & 83.3 \\
 & 10 & 499.25 $\pm$ 23.27 & 13617 $\pm$ 71 & 22299 (0.13\%) & 1947 $\pm$ 152 & 76.7 \\
\midrule
\multirow{6}{*}{10k}
 & 1  & 43.11 $\pm$ 0.92   & 2617 $\pm$ 18  & 27217 (0.16\%) & 1979 $\pm$ 160 & 63.3 \\
 & 2  & 67.59 $\pm$ 2.02   & 3477 $\pm$ 27  & 27003 (0.16\%) & 1967 $\pm$ 161 & 70.0 \\
 & 3  & 100.36 $\pm$ 4.81  & 4718 $\pm$ 19  & 27304 (0.16\%) & 1979 $\pm$ 160 & 63.3 \\
 & 5  & 194.80 $\pm$ 12.02 & 7245 $\pm$ 44  & 27812 (0.17\%) & 1998 $\pm$ 156 & 53.3 \\
 & 7  & 291.46 $\pm$ 9.88  & 9780 $\pm$ 64  & 27915 (0.17\%) & 1973 $\pm$ 160 & 66.7 \\
 & 10 & 290.21 $\pm$ 7.12  & 13614 $\pm$ 71 & 27983 (0.17\%) & 1979 $\pm$ 160 & 63.3 \\
\midrule
\multirow{6}{*}{200k}
 & 1  & 117.96 $\pm$ 4.75  & 2615 $\pm$ 26  & 31288 (0.19\%) & 2017 $\pm$ 150 & 43.3 \\
 & 2  & 128.23 $\pm$ 3.55  & 3481 $\pm$ 32  & 31436 (0.19\%) & 2004 $\pm$ 154 & 50.0 \\
 & 3  & 160.60 $\pm$ 4.82  & 4721 $\pm$ 23  & 31558 (0.19\%) & 2023 $\pm$ 148 & 40.0 \\
 & 5  & 236.90 $\pm$ 6.86  & 7246 $\pm$ 40  & 31785 (0.19\%) & 2023 $\pm$ 148 & 40.0 \\
 & 7  & 399.43 $\pm$ 10.49 & 9790 $\pm$ 47  & 31854 (0.19\%) & 2016 $\pm$ 150 & 43.3 \\
 & 10 & 540.15 $\pm$ 20.12 & 13613 $\pm$ 48 & 31854 (0.19\%) & 2017 $\pm$ 150 & 43.3 \\
\bottomrule
\end{tabular}
\end{table}

\begin{table}[!htbp]
\centering
\caption{IGDec-QAOA recovery of JUnit 4.13.2 with an active set of
9 qubits (3 weight matrices), using the Qiskit AER simulator with the
IBM Montreal noise model (30 runs per configuration).}
\label{tab:igdec-9-montreal}
\footnotesize
\setlength{\tabcolsep}{3pt}
\begin{tabular}{@{}rrccccc@{}}
\toprule
Shots & Rep. & Time (s) & Iter. & States eval. & Avg. obj. & Opt. (\%) \\
\midrule
\multirow{6}{*}{2k}
 & 1  & 1132.06 $\pm$ 41.84    & 2622 $\pm$ 28   & 24142 (0.14\%) & 1960 $\pm$ 161 & 73.3 \\
 & 2  & 1744.64 $\pm$ 73.29    & 3482 $\pm$ 30   & 25140 (0.15\%) & 1973 $\pm$ 160 & 63.3 \\
 & 3  & 2896.57 $\pm$ 163.41   & 4733 $\pm$ 36   & 26002 (0.15\%) & 1992 $\pm$ 158 & 56.7 \\
 & 5  & 5736.52 $\pm$ 417.49   & 7283 $\pm$ 41   & 27707 (0.17\%) & 1992 $\pm$ 158 & 56.7 \\
 & 7  & 9296.20 $\pm$ 699.33   & 9861 $\pm$ 72   & 28710 (0.17\%) & 1992 $\pm$ 158 & 56.7 \\
 & 10 & 17458.82 $\pm$ 1778.09 & 13880 $\pm$ 111 & 29614 (0.18\%) & 2004 $\pm$ 154 & 50.0 \\
\midrule
\multirow{6}{*}{10k}
 & 1  & 1126.85 $\pm$ 40.34    & 2620 $\pm$ 15   & 29620 (0.18\%) & 2017 $\pm$ 150 & 43.3 \\
 & 2  & 1846.42 $\pm$ 134.28   & 3487 $\pm$ 34   & 30307 (0.18\%) & 2023 $\pm$ 148 & 40.0 \\
 & 3  & 2887.78 $\pm$ 190.20   & 4727 $\pm$ 29   & 30904 (0.18\%) & 2023 $\pm$ 148 & 40.0 \\
 & 5  & 5898.53 $\pm$ 762.28   & 7283 $\pm$ 38   & 31794 (0.19\%) & 2023 $\pm$ 148 & 40.0 \\
 & 7  & 9284.01 $\pm$ 835.38   & 9887 $\pm$ 71   & 32180 (0.19\%) & 2023 $\pm$ 148 & 40.0 \\
 & 10 & 17517.82 $\pm$ 1786.97 & 13896 $\pm$ 105 & 32473 (0.19\%) & 2023 $\pm$ 148 & 40.0 \\
\midrule
\multirow{6}{*}{200k}
 & 1  & 1191.19 $\pm$ 61.45    & 2619 $\pm$ 24   & 32223 (0.19\%) & 2023 $\pm$ 170 & 40.0 \\
 & 2  & 1740.84 $\pm$ 84.39    & 3482 $\pm$ 29   & 32460 (0.19\%) & 2023 $\pm$ 148 & 40.0 \\
 & 3  & 2908.04 $\pm$ 170.14   & 4740 $\pm$ 29   & 32582 (0.19\%) & 2023 $\pm$ 148 & 40.0 \\
 & 5  & 5804.44 $\pm$ 485.68   & 7294 $\pm$ 59   & 32632 (0.19\%) & 2023 $\pm$ 148 & 40.0 \\
 & 7  & 9574.17 $\pm$ 886.19   & 9875 $\pm$ 53   & 32639 (0.19\%) & 2023 $\pm$ 148 & 40.0 \\
 & 10 & 17485.19 $\pm$ 1787.42 & 13889 $\pm$ 123 & 32640 (0.19\%) & 2023 $\pm$ 148 & 40.0 \\
\bottomrule
\end{tabular}
\end{table}

\begin{table}[!htbp]
\centering
\caption{IGDec-QAOA recovery of JUnit 4.13.2 with an active set of
8 qubits (4 weight matrices), using the Qiskit AER ideal simulator
(30 runs per configuration).}
\label{tab:igdec-8-ideal}
\footnotesize
\setlength{\tabcolsep}{3pt}
\begin{tabular}{@{}rrccccc@{}}
\toprule
Shots & Rep. & Time (s) & Iter. & States eval. & Avg. obj. & Opt. (\%) \\
\midrule
\multirow{6}{*}{2k}
 & 1  & 61.81 $\pm$ 1.97   & 3505 $\pm$ 21  & 23916 (0.14\%) & 1998 $\pm$ 178 & 60.0 \\
 & 2  & 91.41 $\pm$ 1.60   & 4623 $\pm$ 42  & 23460 (0.14\%) & 1993 $\pm$ 170 & 66.7 \\
 & 3  & 146.08 $\pm$ 3.45  & 6303 $\pm$ 25  & 23730 (0.14\%) & 1979 $\pm$ 174 & 70.0 \\
 & 5  & 289.86 $\pm$ 16.17 & 9677 $\pm$ 45  & 24393 (0.15\%) & 2005 $\pm$ 182 & 60.0 \\
 & 7  & 465.31 $\pm$ 22.11 & 13053 $\pm$ 63 & 24895 (0.15\%) & 2017 $\pm$ 179 & 53.3 \\
 & 10 & 784.12 $\pm$ 22.16 & 18174 $\pm$ 97 & 24927 (0.15\%) & 2016 $\pm$ 186 & 60.0 \\
\midrule
\multirow{6}{*}{10k}
 & 1  & 71.66 $\pm$ 2.38   & 3514 $\pm$ 29  & 28100 (0.17\%) & 2035 $\pm$ 179 & 50.0 \\
 & 2  & 105.61 $\pm$ 4.74  & 4632 $\pm$ 34  & 28019 (0.17\%) & 2035 $\pm$ 179 & 50.0 \\
 & 3  & 184.56 $\pm$ 6.08  & 6321 $\pm$ 32  & 28119 (0.17\%) & 2016 $\pm$ 179 & 56.7 \\
 & 5  & 267.72 $\pm$ 9.29  & 9688 $\pm$ 47  & 28649 (0.17\%) & 2035 $\pm$ 179 & 50.0 \\
 & 7  & 412.22 $\pm$ 13.29 & 13050 $\pm$ 62 & 28845 (0.17\%) & 2029 $\pm$ 182 & 53.3 \\
 & 10 & 861.32 $\pm$ 45.05 & 18145 $\pm$ 90 & 29033 (0.17\%) & 2035 $\pm$ 179 & 50.0 \\
\midrule
\multirow{6}{*}{200k}
 & 1  & 146.88 $\pm$ 2.99  & 3514 $\pm$ 25  & 30245 (0.18\%) & 2035 $\pm$ 179 & 50.0 \\
 & 2  & 173.87 $\pm$ 3.63  & 4624 $\pm$ 38  & 30293 (0.18\%) & 2035 $\pm$ 179 & 50.0 \\
 & 3  & 227.77 $\pm$ 8.18  & 6323 $\pm$ 32  & 30375 (0.18\%) & 2035 $\pm$ 179 & 50.0 \\
 & 5  & 351.02 $\pm$ 9.58  & 9710 $\pm$ 49  & 30503 (0.18\%) & 2035 $\pm$ 179 & 50.0 \\
 & 7  & 529.88 $\pm$ 19.49 & 13044 $\pm$ 79 & 30556 (0.18\%) & 2035 $\pm$ 179 & 50.0 \\
 & 10 & 872.05 $\pm$ 32.05 & 18104 $\pm$ 85 & 30573 (0.18\%) & 2035 $\pm$ 179 & 50.0 \\
\bottomrule
\end{tabular}
\end{table}

\begin{table}[!htbp]
\centering
\caption{IGDec-QAOA recovery of JUnit 4.13.2 with an active set of
8 qubits (4 weight matrices), using the Qiskit AER simulator with the
IBM Montreal noise model (30 runs per configuration).}
\label{tab:igdec-8-montreal}
\footnotesize
\setlength{\tabcolsep}{3pt}
\begin{tabular}{@{}rrccccc@{}}
\toprule
Shots & Rep. & Time (s) & Iter. & States eval. & Avg. obj. & Opt. (\%) \\
\midrule
\multirow{6}{*}{2k}
 & 1  & 1322.47 $\pm$ 58.75   & 3514 $\pm$ 25   & 27008 (0.16\%) & 2029 $\pm$ 182 & 53.3 \\
 & 2  & 1924.12 $\pm$ 67.50   & 4622 $\pm$ 34   & 27810 (0.17\%) & 2035 $\pm$ 179 & 50.0 \\
 & 3  & 2947.38 $\pm$ 78.02   & 6341 $\pm$ 31   & 28600 (0.17\%) & 2035 $\pm$ 179 & 50.0 \\
 & 5  & 5127.81 $\pm$ 168.16  & 9754 $\pm$ 50   & 29651 (0.18\%) & 2035 $\pm$ 179 & 50.0 \\
 & 7  & 7889.32 $\pm$ 271.53  & 13290 $\pm$ 89  & 30203 (0.18\%) & 2035 $\pm$ 179 & 50.0 \\
 & 10 & 13985.45 $\pm$ 724.32 & 18865 $\pm$ 161 & 30512 (0.18\%) & 2035 $\pm$ 179 & 50.0 \\
\midrule
\multirow{6}{*}{10k}
 & 1  & 1228.60 $\pm$ 23.27   & 3515 $\pm$ 31   & 29786 (0.18\%) & 2035 $\pm$ 179 & 50.0 \\
 & 2  & 1882.24 $\pm$ 40.59   & 4632 $\pm$ 33   & 30091 (0.18\%) & 2035 $\pm$ 179 & 50.0 \\
 & 3  & 2835.26 $\pm$ 98.70   & 6327 $\pm$ 31   & 30387 (0.18\%) & 2035 $\pm$ 179 & 50.0 \\
 & 5  & 5216.28 $\pm$ 148.28  & 9774 $\pm$ 57   & 30645 (0.18\%) & 2035 $\pm$ 179 & 50.0 \\
 & 7  & 8062.09 $\pm$ 387.55  & 13253 $\pm$ 82  & 30704 (0.18\%) & 2035 $\pm$ 179 & 50.0 \\
 & 10 & 13800.47 $\pm$ 583.99 & 18877 $\pm$ 149 & 30719 (0.18\%) & 2035 $\pm$ 179 & 50.0 \\
\midrule
\multirow{6}{*}{200k}
 & 1  & 1333.89 $\pm$ 22.26   & 3518 $\pm$ 28   & 30646 (0.18\%) & 2035 $\pm$ 179 & 50.0 \\
 & 2  & 1985.95 $\pm$ 75.99   & 4633 $\pm$ 40   & 30704 (0.18\%) & 2035 $\pm$ 179 & 50.0 \\
 & 3  & 2953.54 $\pm$ 72.52   & 6340 $\pm$ 46   & 30713 (0.18\%) & 2035 $\pm$ 179 & 50.0 \\
 & 5  & 5156.91 $\pm$ 260.37  & 9767 $\pm$ 47   & 30720 (0.18\%) & 2035 $\pm$ 179 & 50.0 \\
 & 7  & 8267.52 $\pm$ 328.51  & 13306 $\pm$ 92  & 30720 (0.18\%) & 2035 $\pm$ 179 & 50.0 \\
 & 10 & 14449.56 $\pm$ 651.33 & 18860 $\pm$ 153 & 30720 (0.18\%) & 2035 $\pm$ 179 & 50.0 \\
\bottomrule
\end{tabular}
\end{table}

\begin{table}[!htbp]
\centering
\caption{IGDec-QAOA recovery of JUnit 4.13.2 with an active set of
12 qubits (4 weight matrices), using the Qiskit AER ideal simulator
(30 runs per configuration).}
\label{tab:igdec-12-ideal}
\footnotesize
\setlength{\tabcolsep}{3pt}
\begin{tabular}{@{}rrccccc@{}}
\toprule
Shots & Rep. & Time (s) & Iter. & States eval. & Avg. obj. & Opt. (\%) \\
\midrule
\multirow{6}{*}{2k}
 & 1  & 70.17 $\pm$ 1.29   & 2634 $\pm$ 18  & 67354 (0.40\%)  & 1867 $\pm$ 1   & 93.3 \\
 & 2  & 111.50 $\pm$ 4.83  & 3467 $\pm$ 33  & 65770 (0.39\%)  & 1898 $\pm$ 97  & 76.7 \\
 & 3  & 187.68 $\pm$ 5.44  & 4749 $\pm$ 28  & 67578 (0.40\%)  & 1880 $\pm$ 68  & 90.0 \\
 & 5  & 383.72 $\pm$ 9.13  & 7285 $\pm$ 39  & 70958 (0.42\%)  & 1892 $\pm$ 95  & 83.3 \\
 & 7  & 579.03 $\pm$ 8.62  & 9819 $\pm$ 59  & 73650 (0.44\%)  & 1873 $\pm$ 17  & 80.0 \\
 & 10 & 994.57 $\pm$ 27.97 & 13609 $\pm$ 68 & 74566 (0.44\%)  & 1880 $\pm$ 68  & 93.3 \\
\midrule
\multirow{6}{*}{10k}
 & 1  & 71.64 $\pm$ 1.09   & 2642 $\pm$ 26  & 140063 (0.83\%) & 1942 $\pm$ 152 & 80.0 \\
 & 2  & 114.63 $\pm$ 1.96  & 3466 $\pm$ 25  & 133077 (0.79\%) & 1954 $\pm$ 160 & 76.7 \\
 & 3  & 180.47 $\pm$ 3.06  & 4731 $\pm$ 29  & 135297 (0.81\%) & 1941 $\pm$ 152 & 80.0 \\
 & 5  & 335.08 $\pm$ 7.69  & 7296 $\pm$ 48  & 144184 (0.86\%) & 1929 $\pm$ 141 & 80.0 \\
 & 7  & 561.55 $\pm$ 11.80 & 9813 $\pm$ 45  & 148907 (0.89\%) & 1942 $\pm$ 151 & 80.0 \\
 & 10 & 985.09 $\pm$ 20.78 & 13600 $\pm$ 70 & 151996 (0.91\%) & 1954 $\pm$ 160 & 76.7 \\
\midrule
\multirow{6}{*}{200k}
 & 1  & 138.44 $\pm$ 3.74  & 2631 $\pm$ 21  & 220229 (1.31\%) & 1973 $\pm$ 160 & 66.7 \\
 & 2  & 171.22 $\pm$ 4.15  & 3461 $\pm$ 28  & 216502 (1.29\%) & 1979 $\pm$ 160 & 63.3 \\
 & 3  & 256.11 $\pm$ 8.04  & 4750 $\pm$ 27  & 219975 (1.31\%) & 1960 $\pm$ 160 & 73.3 \\
 & 5  & 371.49 $\pm$ 11.50 & 7287 $\pm$ 43  & 225697 (1.35\%) & 1985 $\pm$ 159 & 60.0 \\
 & 7  & 619.81 $\pm$ 16.14 & 9810 $\pm$ 69  & 229196 (1.37\%) & 1973 $\pm$ 160 & 66.7 \\
 & 10 & 972.06 $\pm$ 36.09 & 13600 $\pm$ 91 & 231323 (1.38\%) & 1985 $\pm$ 159 & 60.0 \\
\bottomrule
\end{tabular}
\end{table}

\begin{table}[!htbp]
\centering
\caption{IGDec-QAOA recovery of JUnit 4.13.2 with an active set of
12 qubits (4 weight matrices), using the Qiskit AER simulator with the
IBM Montreal noise model (30 runs per configuration).}
\label{tab:igdec-12-montreal}
\footnotesize
\setlength{\tabcolsep}{3pt}
\begin{tabular}{@{}rrccccc@{}}
\toprule
Shots & Rep. & Time (s) & Iter. & States eval. & Avg. obj. & Opt. (\%) \\
\midrule
\multirow{2}{*}{2k}
 & 1 & 11576.85 $\pm$ 250.93  & 2640 $\pm$ 24 & 75630 (0.45\%)  & 1879 $\pm$ 68  & 96.7 \\
 & 2 & 24380.06 $\pm$ 249.92  & 3467 $\pm$ 31 & 77301 (0.46\%)  & 1904 $\pm$ 114 & 90.0 \\
\midrule
\multirow{2}{*}{10k}
 & 1 & 12411.57 $\pm$ 911.51  & 2645 $\pm$ 23 & 160775 (0.96\%) & 1966 $\pm$ 168 & 73.3 \\
 & 2 & 24393.96 $\pm$ 512.18  & 3478 $\pm$ 26 & 164523 (0.98\%) & 1954 $\pm$ 160 & 76.7 \\
\midrule
\multirow{2}{*}{200k}
 & 1 & 12037.78 $\pm$ 603.32  & 2634 $\pm$ 19 & 235900 (1.41\%) & 1992 $\pm$ 158 & 56.7 \\
 & 2 & 24827.26 $\pm$ 1205.16 & 3472 $\pm$ 29 & 241015 (1.44\%) & 1992 $\pm$ 158 & 56.7 \\
\bottomrule
\end{tabular}
\end{table}

\section{Discussion}
\label{sec:discussion}

\paragraph{Effectiveness of QAOA for LAR.}
Monolithic QAOA is feasible for layered architecture recovery and, on small or degenerate instances, matches the classical optimum. Its accuracy, however, drops sharply with the qubit count: usable on $15$ and $18$ qubits, it cannot reliably recover the optimum at $24$ qubits within a realistic shot budget. It is also slow: simulation time grows steeply with the number of qubits, so even a small instance ($8$ packages, $3$ layers) takes a long time to run, while the classical baseline solves the same instance in a fraction of a second. This matches the broader picture in quantum optimisation for software engineering, where QAOA is competitive with but not superior to classical search at present scales~\cite{wang2024,trovato2025}.

\paragraph{Decomposition lifts the qubit ceiling.}
This limitation is one of circuit \emph{width}, not of the formulation. Impact-guided decomposition restricts each circuit to a small active set of packages, and recovers the $24$-qubit JUnit instance---where monolithic QAOA reaches the optimum in at most $6.7\%$ of runs---in up to $93\%$ of runs (RQ4). Because the optimal cost is the same ($1867$) at three and four layers, the two rates measure the same target and the improvement is direct.

\paragraph{Two different search mechanisms.}
The two pipelines succeed in different ways. In the monolithic pipeline the real search is done by classical post-processing: solution quality follows the fraction of the $2^{mn}$ space it evaluates, which is high on the small systems ($60$--$88\%$) and collapses to about $5.6\%$ on JUnit. Under decomposition this no longer applies. Each sub-problem is small enough to be solved almost exactly, so the quantum circuit is not the bottleneck; what matters is how the greedy outer loop sweeps the packages. This is why a larger active set helps, and why more shots make the result \emph{worse}: a large shot budget makes every step strictly greedy and easy to trap, while a small budget adds the sampling noise that lets the sweep escape poor solutions.

\paragraph{Cost and parameter choices.}
In both pipelines, circuit depth ($p$) is expensive and adds little accuracy, so a shallow circuit ($p{=}1$) is the efficient choice. The shot budget, however, should be set differently: the monolithic pipeline needs a large budget scaled to the problem size, whereas decomposition works best with the \emph{smallest} budget, which makes it cheap in both depth and sampling.

\paragraph{Noise and the path to hardware.}
Simulated noise did not degrade solution quality on the smallest monolithic instance, but it raised simulation cost by one to two orders of magnitude and made the $18$- and $24$-qubit monolithic cases infeasible. Decomposition partly lifts this barrier: because each sub-circuit is narrow, the $24$-qubit JUnit instance ran under the IBM Montreal noise model (RQ4) with quality largely preserved, at a time cost of one to two orders of magnitude. Beyond this, progress on larger instances depends less on the QAOA formulation than on quantum error mitigation and eventual execution on real hardware, where the cost profile differs from density-matrix simulation.

\section{Threats to Validity}
\label{sec:threats}

\paragraph{Construct validity.}
The ground-truth layered architectures were obtained partly by manual analysis, following Belle et al.~\cite{belle2015}, and the ``optimal'' against which the optimal rate is measured is the best known assignment for each instance. The best known cost is the same ($1867$) at three and four layers, so the optimal rates of the monolithic and decomposed runs are measured against the same target and are comparable. The LAR cost parameters ($\alpha,\beta,\gamma$) are fixed; different penalties could change which assignment is optimal, and hence the reported rates.

\paragraph{Internal validity.}
QAOA results are stochastic; we mitigate this by averaging over $30$ runs and reporting standard deviations. COBYLA may converge to local optima, and we did not tune its termination tolerance, which could affect runtime and accuracy. As the reported solutions come from quantum sampling followed by classical post-processing, the figures describe the hybrid pipeline, not quantum sampling alone. For the decomposition, the random initial assignment and the greedy, impact-ordered sequence of sub-problems add further randomness and a dependence on ordering that we do not isolate here. The active-set size was also chosen on the same JUnit instance on which results are reported, so the best configuration may be partly specific to that instance.

\paragraph{External validity.}
The study covers three relatively small Java systems, with at most $24$ qubits in the monolithic setting, which limits generalisation to larger systems. The decomposition was run at both three and four layers, but no system uses more layers than this, so how recovery quality scales with the number of layers remains open. Results were obtained on simulators, which may not predict real-hardware behaviour, and the noise models used---IBM Quebec, depolarising, and dephasing for the monolithic runs, and IBM Montreal for the decomposition--- are approximations of device noise.

\paragraph{Conclusion validity.}
The classical baseline uses the Tabu Search configuration of Belle et al.~\cite{belle2015} for comparability; another solver, or a tuned Tabu Search, could shift the comparison. The infeasible noisy runs at $18$ and $24$ qubits reflect our compute budget on a single Narval node, not a fundamental limit. Finally, the decomposition results come from a single system (JUnit), so they should be read as a demonstration of feasibility rather than a general performance claim.

\section{Related Work}
\label{sec:related}

\paragraph{Software architecture recovery and remodularization.}
Recovering or improving a system's modular structure has a long history in search-based software engineering. Anquetil and Lethbridge studied clustering as a remodularization method~\cite{anquetil1999}; Maqbool and Babri surveyed and advanced hierarchical clustering for architecture recovery~\cite{maqbool2007}; and Praditwong, Harman, and Yao framed module clustering as a multi-objective search problem~\cite{praditwong2011}, a line later extended to many-objective remodularization with NSGA-III~\cite{mkaouer2015}. These problems are NP-hard and are typically tackled with metaheuristics. Our work targets the specific layered-recovery formulation of Belle et al.~\cite{belle2015}, who cast it as a quadratic assignment problem and solved it with Tabu Search; we reuse that formulation and baseline and ask whether a quantum optimizer can address it.

\paragraph{Quantum optimization for software engineering.}
Quantum and quantum-inspired optimization is an active and recent thread in software engineering, surveyed by Zhang et al.~\cite{zhang2026survey} and positioned in the broader roadmaps of Wang et al.~\cite{wang2025roadmap}. Most concrete results so far concern testing. Wang et al.\ formulated test case optimization for QAOA with a decomposition strategy for large instances, evaluating it on ideal and noisy simulators and on real hardware~\cite{wang2024}; the same group addressed test case minimization with quantum annealers~\cite{wang2024minimization}; and Trovato et al.\ investigated QAOA for test case selection~\cite{trovato2025}. A recurring finding across this work is that QAOA can reach but rarely exceed classical effectiveness at NISQ scale. This finding matches what we observe for architecture recovery. To our knowledge, this is the first study to apply QAOA to layered architecture recovery, and to characterise its parameter sensitivity, classical comparison, and noise behaviour for that problem.

\section{Conclusion and Future Work}
\label{sec:conclusion}

We presented a hybrid quantum--classical method for layered architecture recovery that formulates the problem as a QUBO, encodes it as an Ising Hamiltonian, and solves it with QAOA followed by classical post-processing, together with an impact-guided decomposition that keeps the circuit width within near-term budgets. An empirical study on three real Java systems showed that monolithic QAOA is feasible and can match the classical optimum on small or degenerate instances, that the sampling budget rather than circuit depth drives its accuracy, that QAOA is competitive with but does not beat Tabu Search at current scales, and that simulated noise preserves quality on the smallest instance while making monolithic simulation infeasible beyond fifteen qubits. Impact-guided decomposition lifts this barrier on the $24$-qubit JUnit instance: by limiting each circuit to a small active set of packages, it raises the optimal rate from $6.7\%$ to $93\%$ and keeps noisy simulation feasible where the monolithic pipeline could not run. Unlike the monolithic pipeline, it works best with a small sampling budget.

These results position QAOA for software architecture tasks as promising but constrained by the NISQ era, with problem decomposition, rather than larger circuits or more sampling, emerging as the key to reaching larger instances. Our future work follows from the limitations identified here: applying quantum error mitigation to make noisy execution tractable; running the pipeline on real quantum hardware, where the cost profile differs from noisy simulation; and extending the evaluation to larger and more varied systems, including a study of how recovery quality scales with the number of layers.

\section*{Acknowledgements}
This work was supported by Mitacs and PINQ$^2$ through the Accelerate Program (IT36314). The authors also thank Calcul Qu\'ebec and the Digital Research Alliance of Canada for providing the computing resources used in this study.


%

\bibliography{references}

\end{document}